\documentclass[useAMS,usenatbib]{mn2e}
\usepackage{graphicx} 

\title[GMRT Low Frequency Observations of Extrasolar Planetary Systems]
{GMRT Low Frequency Observations of Extrasolar Planetary Systems}
\author[S. J. George and I. R. Stevens]{S. J. George$^{1}$
\thanks {E-mail:samuel@star.sr.bham.ac.uk (SJG)} and I. R. Stevens$^{1}$
\footnotemark[0]\\
$^{1}$School of Physics and Astronomy, University of Birmingham, 
Edgbaston, Birmingham, UK B15 2TT}
\begin{document}
\date{Accepted 2007 August 22. Received 2007 July 23; in original form 2007 May 30}

\pagerange{\pageref{firstpage}--\pageref{lastpage}} \pubyear{2007}

\maketitle

\label{firstpage}

\begin{abstract}

Extrasolar planets are expected to emit detectable low frequency radio
emission. In this
paper we present results from new low frequency observations of two
extrasolar planetary systems (Epsilon Eridani and HD\,128311) taken at
150~MHz with the Giant Metrewave Radio Telescope (GMRT). These two
systems have been chosen because the stars are young (with ages $<1$
Gyr) and are likely to have strong stellar winds, which will increase
the expected radio flux. The planets are massive (presumably) gas
giant planets in longer period orbits, and hence will not be tidally
locked to their host star (as is likely to be the case for short
period planets) and we would expect them to have a strong planetary
dynamo and magnetic field. We do not detect either system, but are able to place tight upper
limits on their low frequency radio emission, at levels comparable to
the theoretical predictions for these systems. From these observations
we have a $2.5\sigma$ limit of $7.8$~mJy for $\epsilon$ Eri and
$15.5$~mJy for HD\,128311.  In addition, these upper limits also
provide limits on the low frequency radio emission from the stars
themselves.
These results are discussed and also the prospects for the future detection of radio emission from
extrasolar planets.

\end{abstract}

\begin{keywords}
extrasolar planets -- observations: radio
\end{keywords}

\section{Introduction}
\label{radio_detec}

The first definite extrasolar planet orbiting a normal star was
detected in 1995 (Mayor \& Queloz 1995). The main current techniques
for detecting extrasolar planets (Doppler, transits etc) are indirect
in nature, seeing the influence of the planet on the light from the
host star rather than detecting the planetary properties
directly. These methods infer rather limited information about the
planet (such as the mass, or often just a lower limit on the
mass). Observations at other wavelengths, such as low radio
frequencies, have the potential for determining certain planetary
properties that will be important in understanding planetary
structure and evolution. 

By analogy with the magnetic planets in the solar system, observations
at radio wavelengths of extrasolar planets offers the (as yet
unrealised) potential for a new way of directly detecting extrasolar
planets and also a means of inferring information about the
characteristics of the planets, such as the existence of a
magnetosphere, the planetary magnetic field strength, planetary
rotation and even the presence of moons, that may be difficult via
other means (cf Higgins et al. 1997), and which will be important in
developing our understanding of how these planets form and evolve.

Low frequency radio observations of solar system planets have been
made for decades (see, for example, Bastian et al. 2000).  Burke \&
Franklin (1955) were the first to discover radio emission from
Jupiter, which is an extremely bright and variable source at
decametric (3--30~MHz) wavelengths. Due to ionospheric cut-off, it was not until the
Voyager spacecraft era that the other solar system gas giant planets
were detected (Broadfoot et al. 1981).

For solar system planets, the dominant mechanism for the low
frequency radio emission is believed to be the electron cyclotron
maser. The characteristic emission frequencies are determined by the magnetic 
field strength near the surface of the planet. The level of emission is 
dependent on local conditions (such as rotation and satellite interactions) 
but in essence all that is required is there to be a distribution of energetic electrons 
within the planetary magnetic field, with an anisotropy in velocity space and a large ratio 
of gyro to plasma frequencies.

We expect that massive extrasolar planets will, like Jupiter, have
a dynamo driven magnetic field and a similar interaction with the wind
of their host star. If this is the case then they will produce radio
emission. All the larger planets in our own solar system produce radio
emission, so there is no reason why we should not expect the same to
occur for extrasolar planets (Ness et al. 1986). 
The major differences between extrasolar planets and 
solar system planets will be the sensitivity required to detect them
and the frequencies at which the emission will occur (Stevens 2005).

The radio detection of extrasolar planets has been attempted several
times, with a variety of different telescopes, at different
frequencies. Winglee et al. (1986) used the Very Large Array (VLA) (at 0.33~GHz and 1.4~GHz) to observe
six nearby stars, however no detections were made. 
Bastian et al. (2000) also used the VLA for multi-frequency observations of extrasolar planets at
1.4~GHz, 0.33~GHz and 74~MHz. No detections were reported, with
sensitivities of $\sim 50$~mJy, at the lowest frequency used.
More recently, the Ukrainian UTR-2 telescope has surveyed around 20
extrasolar planetary systems in the 10--25~MHz range, with a
sensitivity of about 1.6~Jy. This survey produced no detections
(Ryabov et al. 2003).
Further to this Lazio \& Farrell (2007) have observed $\tau$ Bootes
with the VLA at 74~MHz. No detection was made, with an upper limit of
around 150~mJy.
Winterhalter et al. (2006) have used the GMRT to observe extrasolar
planets, focusing on short period systems. The targets observed were
$\tau$ Bootes, HD 162020, HD 179949, 70 Virginis and $\upsilon$
Andromedae. No detections were made.

In addition to these pointed observations, several large area radio
surveys exist and these can provide constraints on the emission from
extrasolar planets. These include the Cambridge 6C (150~MHz), 7C
(150~MHz) and 8C (38~MHz) surveys and the VLA Low-Frequency Sky Survey
(VLSS, 74~MHz). For details of these surveys see Hales et al. (1995) and
Lazio et al. (2004).

There may be several reasons for these non-detections; insufficient
sensitivity or incorrect (or even unlucky) target or frequency
selection are probably key. We do not know how the radio emission from
Jupiter-like planets really scales in other systems. There are also
issues of beaming of the emission and time variability.  Nonetheless,
the prize of detecting extrasolar planets at radio wavelengths is
sufficiently great to warrant continued searches.

The paper is organised as follows: in Section~2 we review our
understanding of the expected radio emission from extrasolar
planets. In Section~3 we introduce the GMRT targets and selection rationale. 
The data analysis and results are
presented in Section~4, and in Section~5 we discuss the consequences
and also briefly discuss limits from low-frequency observations.

\section{Low Frequency Planetary Radio Emission}
\label{Jovian-system}
Before we discuss extrasolar planetary radio emission, it will be
useful to review the Jovian low frequency radio emission.
Below a cut-off frequency of $\sim 40$~MHz, the Jovian radio emission
is thought to be dominated by cyclotron-maser processes from keV
electrons in the auroral regions of the planet. The power radiated by Jupiter at this frequency
range is several hundred Gigawatts and below the cut-off frequency the radio power 
is comparable to that of the quiet Sun (Winterhalter et al. 2006). The emission is variable and solar events
such as coronal mass ejections cause the emission intensity to
increase. Apart from the solar wind connection there is an
additional component related to Io. The moon-planet interaction
produces a significant increase in decametric emission (Gurnett et al. 2002).
At higher (GHz) frequencies, synchrotron emission generated from the
acceleration of energetic electrons, trapped in Jupiter's magnetic
field dominates the emission process. This means for detection
of an extrasolar planet the intrinsic planetary emission
will have to be much brighter than Jupiter. This can be achieved by
the planet being in a shorter period orbit than Jupiter, or being more massive than Jupiter, 
or to be orbiting a star with a denser stellar wind than the Sun. 
The first two points have no impact on the radio
flux of the star, but the third will. The stellar mass-loss rate $\dot M_\ast$  will 
affect the planetary radio emission, and is related to the stellar luminosity $L_x$ (Wood et al. 2002).
From the scaling law for radio emission for extrasolar planets, we
expect the planets observed here to be much more luminous than
Jupiter, meaning their expected flux levels could be in the the mJy
range and potentially detectable (Stevens 2005).

Consequently, although Jupiter and the Sun can have comparable
luminosities at low frequencies, for the brighter extrasolar planets,
we expect the planets to be much brighter than their stellar
hosts. This argument assumes constant sources, rather than flaring
sources, and stellar flare activity may boost the stellar emission for
short periods. In addition, it is believed that extrasolar planets will have a highly
polarised emission (due to the emission process) whereas the level of
polarisation for the stellar emission will vary considerably (Le Queau
1987).

Many extrasolar planets are in short period orbits, moving within 
the stellar corona of their host star, and we may expect some complicated 
magnetospheric geometries and interactions in such cases (Ip et al. 2004). 
On the other hand, any short period extrasolar planet may well be tidally locked, with a
possible impact on the planetary dynamo (Grie{\ss}meier et al. 2004).

If we consider the case where the stellar wind is incident on the
planetary magnetosphere then we can produce a scaling relationship
relating the expected radio flux ($P_{r}$) of the planet to the
local physical conditions (for a more detailed discussion see Stevens
2005, Lazio et al. 2004 and Zarka et al. 2001 and references therein), such that

\begin{equation}
P_{r} \propto \dot{M_\ast}^{2/3}V_{w}^{5/3}\mu_{p}^{2/3}A^{-4/3} D^{-2},
\end{equation}
where $\dot{M_\ast}$ is the stellar mass-loss rate, $V_{w}$ is 
the stellar wind velocity incident on the planetary magnetosphere, $\mu_{p}$ 
is the planetary magnetic moment, $A$ is the orbital distance between the 
star and the planet, and $D$ the distance to the Earth.

We might expect the planetary magnetic moment to roughly scale with
the planetary mass. For the solar system, this relation is known as
Blackett's law (Blackett 1947). One form of Blackett's law is 
$\mu_p \propto \omega M_{p}^{5/3}$, where $\mu_p$ is the planetary magnetic moment, $M_p$ the planetary
mass, and $\omega$ the planetary angular rotation frequency. There are other
forms of this relationship, such that $\mu_p \propto M_{p}$
but in general the more massive the planet, the more massive will be
the planetary magnetic moment, resulting in a larger magnetosphere and
enhanced radio emission. Of course, these relations are based on the solar 
system and are not exact. We also have no evidence that they hold for exoplanetary systems 
but they allow us to make promising predictions.

\begin{table*}
\centering
\caption{The stellar and planetary orbital properties of the GMRT
targets. The orbital data for the $\epsilon$ Eri system are from
Benedict et al. (2006) and for HD\,128311 from Vogt et al. (2005).
For $\epsilon$ Eri b the inclination is known ($i=30^\circ.1$), so
that the planetary mass can be determined as $1.55~M_{Jup}$ (Benedict
et al. 2006).  The estimated stellar mass-loss rates are from Stevens
(2005). The separation $A$ is the star-planet separation at the time
of the observations. \label{gmrt_targets1} }\smallskip
\begin{tabular}{lcccccccc}
\hline
   & \multicolumn{3}{c}{Stellar Parameters} & \ &\multicolumn{4}{c}
{Planetary Parameters}\\\cline{2-4}\cline{5-9}
Name & Spectral & Distance & Mass-loss & & Orbital &
   Orbital & Mass & Star-Planet\\ 
     & type &  &  rate & & Eccentricity & period & $M_{p}\sin i$ &  Separation \\ 
     & & $D$ (pc) & $\dot M_\ast$ ($M_{\odot}$) & & $e$ & $P_{orb}$ (days) & ($M_{Jup}$) & $A$ (au)\\
\hline
$\epsilon$ Eri   & K2V & 3.22 &17.4 & & & &&\\
$\epsilon$ Eri\,b&           &      && &0.70&2502&0.78& 2.76 \\ \hline
HD\,128311 & K0V & 16.6 &26.0&&&&&\\ 
HD\,128311\,b &&&&&0.25&458.6&2.18& 1.04\\ 
HD\,128311\,c &&&&&0.17&928.3&3.21& 1.70 \\\hline
\end{tabular}
\end{table*}

\section{Target Selection}
\label{targets}

Based on our knowledge of the situation in the solar system, the level
of radio emission from a given magnetised extrasolar planet will be
proportional to the stellar-wind ram-pressure flux incident on the
planetary magnetosphere. This means that the expected level of radio
emission from a given extrasolar planet will be a function of several
factors, as expressed in eqn.~(1).

The planetary period will greatly affect the level of stellar wind
flux incident on the magnetosphere. Shorter period planets will be
immersed in denser regions of the stellar wind. Although this
compresses the magnetosphere, the net effect is to make the planet
more radio luminous. Observations of extrasolar planets have focused
on the short period systems, because from the scaling laws they are
predicted to be the most luminous (potentially several thousand times
more luminous than Jupiter). However, these systems have not been
detected so far.

Planets that are close to their host star will experience tidal
locking (Grie{\ss}meier et al. 2004). This may cause them to have 
a weak internal magnetic field and any dynamo action may be negligible, 
leading to a small magnetic moment, a reduced magnetosphere and consequently 
less radio emission. This is, of course, speculative and there is evidence for
some magnetic activity associated with short period extrasolar planets
which may indicate that even short period planets have some magnetic
field (Shkolnik et al. 2003).

It is also useful to consider the stellar wind properties of the host
star. Younger, X-ray bright stars, will have a stronger stellar wind
than their older counterparts. From eqn.~(1), we expect that the radio
flux to scale as $\dot M_\ast^{2/3}$, and there will be an advantage
to choosing planets around younger, more X-ray active stars. 
Stevens (2005) assumed that $\dot M_\ast$ scaled
linearly with X-ray luminosity (based on the work of Wood et
al. 2002). The evidence for this scaling has weakened recently, and
may well be weaker (for example, see Cranmer 2007 for a discussion of
this).

\subsection{The GMRT Targets}

A number of previous searches have tended to focus on short period extrasolar
planets, which from simple scaling laws are expected to be the most
luminous (Winterhalter et al. 2006). In
contrast, we have chosen to investigate longer period systems, since
extrasolar planets which are particularly close to their host star
might experience tidal locking (Grie{\ss}meier et al. 2004).

In the absence of any detections we do not know how the radio flux
actually scales. The failure to detect short period systems means that
the strategy of also observing longer period systems is perhaps
prudent, just in case the short period systems turn out to be faint
radio emitters.This strategy does have the disadvantage that the stellar wind density
incident on the planetary magnetosphere declines sharply with orbital
period. This effect is seen in our own solar system where planets with
larger magnetic moments do not necessarily produce the most radio
emission; Uranus has a larger magnetic moment than the Earth but
produces less radio emission (Bergstralh 1987).

Using the criteria discussed above we have selected 2 known extrasolar
planetary systems to observe with the GMRT at 150~MHz; $\epsilon$
Eridani and HD\,128311. Some of the relevant parameters of these
systems are shown in Table \ref{gmrt_targets1}, and the details of
each system discussed below.

\begin{figure*}
\includegraphics[scale=0.45]{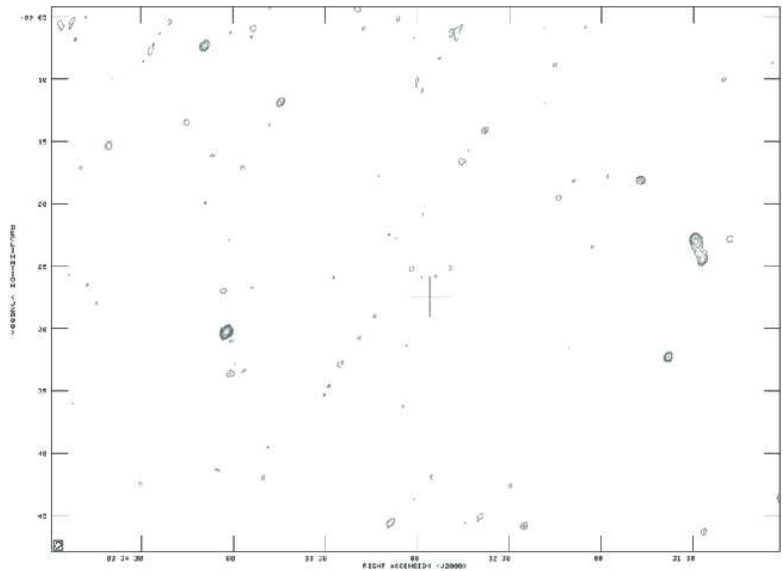}
\caption{The 150~MHz view of the region around $\epsilon$ Eri
(J2000.0 position marked with a cross). No emission is detected from 
$\epsilon$ Eri. The contours levels are 12.5, 25, 50, 100, 200, 400, 800 and 
1600 mJy. The brightest source in the field is NVSS J033322-084803
($\alpha=03^h33^m22.80^s$, $\delta=-08^d48^m01.0^s$) with a flux of 2.85~Jy.}
\label{eps_eri_region}
\end{figure*}

\subsection{Epsilon Eridani}

$\epsilon$ Eri is one of the nearest solar-type stars, at a distance
of 3.22~pc. It is classified as K2V, and is slightly metal-poor
(Fe/H$=-0.13\pm 0.04$; Santos et al. 2004). Due to the high level of
chromospheric activity observed radial velocity variations have been
suspected as arising from the stellar activity cycle. $\epsilon$ Eri
has been a target of many searches for planetary companions. Hubble Space
Telescope astrometric observations have confirmed the presence of a
planet (Benedict et al. 2006). $\epsilon$ Eri~b is a moderately long
period, highly eccentric extrasolar planet ($P_{orb}=2502$ days, 
semi-major axis = $3.39$~au, and
$e=0.70$). This planet is possibly not the only planetary companion in
the $\epsilon$ Eri system.  Observations in the infrared (using the
IRAS satellite) have been used to suggest the presence of planets at
large orbital radii (Quillen \& Thorndike 2002). The debris disk that
surround $\epsilon$ Eri suggest relative youth in the system (0.66
Gyr; Saffe et al. 2005). SCUBA measurements have suggested that there
is a distributed ring, at an orbital radius of about 65~au, around
$\epsilon$ Eri (Greaves et al. 1998) though there has been no clear
evidence of a planet from optical observations. We concentrate on the planet 
discussed by Benedict et al. (2006).

$\epsilon$ Eri is an active star with an X-ray luminosity, of
$L_X=2.1\times 10^{28}$~erg~s$^{-1}$. According to the scaling of Wood
et al. (2002) this results in an estimated mass-loss rate of $\dot
M_\ast =17 \dot M_\odot$. This probably should be regarded as an upper
limit, as recent work suggests a more gradual scaling with $L_X$
(Cranmer 2007). The high eccentricity of
$\epsilon$ Eri~b means that the radio flux is predicted to change by a
factor 3.5 over the course of the $\sim 7$ year orbit. The last periastron
passage occurred at 2007.29 and the next apastron passage at 2010.71
(Benedict et al. 2006). Our observations were taken at 2006.8 which
means that the star planet separation was equal to about 2.76~au. The
inclination of the planet orbit is known ($i=30^\circ.1\pm 3^\circ.8$),
which, combined with the radial velocity solution gives a planetary
mass of $M_p=1.55\pm 0.24 M_{Jup}$.

\subsection{HD\,128311}

HD\,128311 is young (age of $0.5-1.0$ Gyr) active K0 star (Vogt et al. 2005). 
HD\,128311 is estimated to be 16.6~pc
away from the Earth. It has an X-ray luminosity of $L_X=3.0\times 10^{28}$~erg~s$^{-1}$ which 
results in an estimated mass-loss rate of $\dot M_\ast = 26 \dot M_\odot$.
There are two planets in orbit around HD\,128311 with orbital periods
of 458.6 and 928.3 days, with semi-major axes of 1.10~au and 1.76~au 
respectively (Vogt et al. 2005). These planets are quite
eccentric and this has implications to the solar wind flux
encountered. The effect will not be as pronounced as that in
$\epsilon$ Eri b. Using the Vogt et al. (2005) orbital solution, the
star-planet separation was about 1.04~au for HD\,128311\,b and 1.70~au for
HD\,128311\,c at the time of observation.

\begin{figure*}
\includegraphics[scale=0.45]{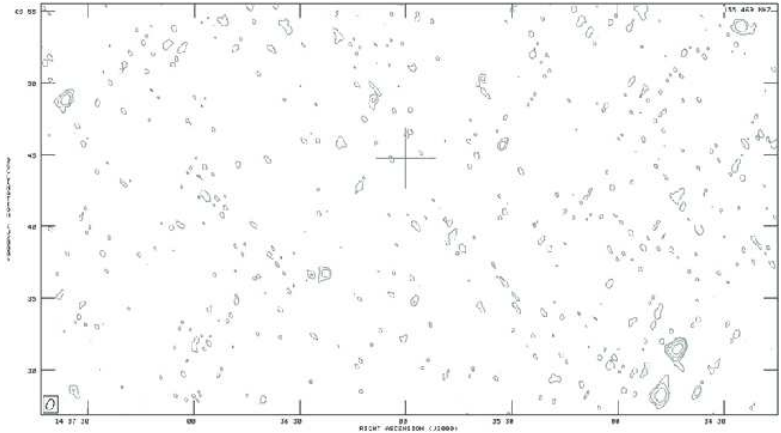}
\caption{The 150~MHz view of the region around HD\,128311 (J2000.0
position marked with a cross). No emission is detected from
HD\,128311. The contours levels are 25, 50, 100, 200 and 400 mJy. The brightest source in the field is TXS 1432+07 ($\alpha=14^h34^m43.25^s$, $\delta=-09^d32^m45.64^s$) with a flux of 0.45~Jy.}
\label{HD128311_region}
\end{figure*}

\section{Data Analysis and Results}

The radio observations of $\epsilon$ Eri and HD\,128311 were conducted
with the GMRT, during cycle 11 on August 10-11 2006, for a duration of
4.13 hours each. The central frequency of the
observation was 157.0~MHz and a bandwidth of 8~MHz was used.
For $\epsilon$ Eri, 3C48 was used as the flux calibration and
$0114-211$ was used as the phase calibration with a time on source in
a single scan of 30 minutes before moving to the phase calibration for
5 minutes. For HD\,128311, 3C286 was used for flux calibration and
$1419+064$ was used as the phase calibration, the same procedure as
for $\epsilon$ Eri was followed.

The data was calibrated using standard procedures in AIPS 
(Astronomical Image Processing System).
Unfortunately at this frequency, the data is significantly affected by
RFI, which was identified and excised manually using AIPS procedures.
Wide field mapping with 5 degree diameter (close to full width of the beam)
consisting of 121 facets were performed to correct for curvature of the sky.
Several iterations of phase-only self calibration was applied to correct for 
short term phase changes. After phase-only self calibration has converged, 
one round of amplitude and phase self calibration was carried out.  
The 121 facets were then combined to make one final image. The final 
restoring beam (resolution) is $32.1'' \times 23.2''$ for $\epsilon$ Eri and 
$45.4'' \times 30.6''$ for HD\,128311. While these stars have substantial proper motions ($\sim 1000$~mas/yr
for $\epsilon$ Eri, and 320 mas/yr for HD\,128311, Perryman et
al. 1997), this has no impact on the astrometry of these observations.

\subsection{Results and Flux Limits}

In the full field of view (of area $\sim 10^\circ \times 10^\circ$) 
for both the $\epsilon$ Eri and HD\,128311 fields we have detected a 
large number of sources. 
For the two observations the RMS noise levels are $\sigma=3.16$~mJy
($\epsilon$ Eri) and $\sigma=6.20$~mJy (HD\,128311). We define a source as one that
has a flux of at least $5\sigma$.
In the $\epsilon$ Eri field, a total of 571 sources were
detected with flux levels between 16~mJy and 2.85~Jy. For HD\,128311 the
corresponding numbers are 297 sources, with fluxes between 30~mJy and
0.45~Jy. These sources are most likely to be background AGN but
at 150~MHz one might expect to detect the presence of a 
starburst galaxy population, as starburst galaxies have a 
steeper spectrum than many AGN. Moss et al. (2007) have 
used the GMRT to undertake a 610~MHz survey to detect such objects. 
We expect that lower frequency observations will assist in the detection 
of these sources and this will be discussed elsewhere (George, in prep).
Images of the smaller regions  around
$\epsilon$ Eri and HD\,128311 are shown in Fig.~\ref{eps_eri_region}
and Fig.~\ref{HD128311_region} respectively. The locations of
$\epsilon$ Eri and HD\,128311 are marked with crosses in the
respective diagrams.

In Fig.~\ref{eps_eri_region}, there are two bright sources in the
vicinity of $\epsilon$ Eri. These are NVSS J033322-084803
($\alpha=03^h33^m22.80^s$, $\delta=-08^d48^m01.0^s$), which has a
150~MHZ flux of 2.85~Jy, and PMN J0331-0923 ($\alpha=03^h31^m25.8^s$,
$\delta=-09^d23^m58^s$) which is a double lobed source.

In Fig.~\ref{HD128311_region}, the brightest source is at
 $\alpha=14^h34^m43.25^s$, $\delta=-09^d32^m45.64^s$, most likely TXS 1432+07,
and has a 150~MHz flux of 0.45~Jy.

It is clear from both Fig.~\ref{eps_eri_region} and
Fig.~\ref{HD128311_region} that neither target source has been
detected at 150~MHz, and we focus on the upper limits on the radio
emission that can be derived from these observations. We have adopted
the same approaches as Lazio \& Farrell (2007) to derive the
flux limits for the extrasolar planets.
We take two different approaches to calculating the flux limit for
each source. The first estimate is from the RMS noise level ($\sigma$)
for each observation, and we adopt the flux upper limit of
2.5$\sigma$. Using this method we have a flux limit of 7.9~mJy 
for $\epsilon$ Eri and 15.5~mJy for HD\,128311.
We also use the brightest pixel within a beam centered on
the targets to estimate the flux density of any possible radio emission 
and allows us to take into account eh background level determined by the 
mean brightness in a region surrounding the central beam. We are determine 
upper limits via this method of 13.8~mJy for $\epsilon$ Eri and 
19.4~mJy for HD\,12811. These upper limits are summarised in Table~\ref{results_table}.

\begin{table}
 \centering
\caption{The derived flux limits for the extrasolar planetary systems observed with the GMRT at 150~MHz. \label{results_table}}\smallskip
\begin{tabular}{|l|c|c|}
\hline
Planet Name & Flux limit ($2.5\sigma$) & Brightest Pixel\\
            & (mJy) & (mJy)\\
\hline
$\epsilon$ Eri b   &  7.9 &  13.8 \\
HD\,128311 b       &  15.5 &  19.4 \\
\hline
\end{tabular}
\end{table}

\section{Discussion}
\label{Discussion}

These GMRT observations have provided tight upper limits on the
150~MHz flux from two nearby, longer period extrasolar planets. In
addition to these observations we can also add in VLSS 74~MHz limits for
these two objects. For $\epsilon$ Eri this is approximately 0.9~Jy and 1.2~Jy 
for HD\,128311. Previous searches for extrasolar planets have focused on short period
systems, but have also failed to detect anything. These observations
complement and extend the range of extrasolar planetary systems that
have been surveyed.

Our observations do
not quite reach the expected radio flux of these systems. Stevens
(2005) suggests that the mean radio flux for $\epsilon$ Eri b would be
2.4~mJy and 1.4~mJy for HD\,128311 b, of course these are only mean
values and do not take account for the eccentricity of the system. In
both cases we have observed the planet closer than at its mean
position so we would expect the flux to increase. Our flux limits do
approach these values. There are several reasons why we have not detected these systems. 

\begin{enumerate}
\item Insufficient sensitivity: One issue for longer period periods
  systems is that they are predicted to be intrinsically fainter than
  the shorter period systems, with the expected flux falling off as
  $P_{orb}^{-8/9}$ (all other things being equal). Clearly more
  sensitive observations are required.

\item Wrong frequency: The radio cut-off for Jupiter is around 40~MHz,
  and it could be the case that 150~MHz is too high a frequency to
  detect the electron-cyclotron maser emission. The Jovian cut-off is
  determined by the magnetic field strength near the planetary surface,
  and probably scales roughly linearly with planetary
  mass. Consequently, we might expect to only detect planets with a
  mass of $M_p\geq 3.8 M_{Jup}$. The mass of $\epsilon$~Eri~b
  ($M_p=1.55 M_{Jup}$) is well constrained and is below this
  value. For HD\,128311, we have only lower limits for the planetary
  masses (of 2.2 and $3.2 M_{Jup}$). If the system inclination is less
  than $\sim 60^\circ$ then one of the planets will have a mass
  greater than this limit and if $i<35^\circ$ (and the planets are
  co-planar) then both will exceed this limit. While not guaranteed,
  there is a moderate chance that 150~MHz is an appropriate frequency
  for detection. Clearly, simultaneous observations in a much wider
  frequency range is desirable, particularly at lower frequencies.

\item Wrong planetary orientation: In the case of the solar system
  planets the low frequency emission comes from a hollow cone with a
  wide opening angle ($\sim 75^\circ$ half-width opening angle for
  Jupiter). For Jupiter the emission has an equivalent
  solid angle of 1.6 sterad (Zarka \& Cecconi 2004). Assuming this 
  also applies to extrasolar planets then we would
  expect to see emission from a given extrasolar planet at some
  epoch. However, this does raise the related issue of time
  variability. It could be that we simply observed the systems at the
  wrong time.
\end{enumerate}

To solve the problem of detecting the magnetospheric emission from
extrasolar planets we can look forward to the Low Frequency Array
(LOFAR). LOFAR is expected to operate in the frequency range of
10-240~MHz. The design goals suggest that a sensitivity, in a 15 minute 
integration with a 4~MHz bandwidth, of around 2~mJy will possible.  
Several features of LOFAR make it a highly capable instrument for detecting 
extrasolar planets --first, it has an appropriate frequency coverage, second it is more
sensitive than current telescopes and thirdly its design means that it
can survey large portions of the sky. In spite of our theoretical
prejudices, we really do not know which extrasolar planetary systems
will turn out to be brightest, and the wide field capability of LOFAR
will be more productive than the repeated pointed observations that
have hitherto been done. We expect that extrasolar planetary radio emission will be
detectable with LOFAR and comparable arrays, if produced in a manner
similar to the solar system planets.

In summary, we have presented GMRT observations of two extrasolar
planetary systems ($\epsilon$ Eri and HD\,128311) at 150~MHz. We have
not detected either systems, but have provided tight upper limits on
the low-frequency radio emission from the magnetospheres of these planets.

\section*{Acknowledgements}

We thank the staff of the GMRT who have made these observations possible. 
The GMRT is run by the National Centre for Radio Astrophysics of the Tata 
Institute of Fundamental Research. In particular we would like to thank 
Ishwara Chandra C. H. for his input during the observations and with 
the data analysis.

\noindent SJG is supported by a PPARC/STFC studentship.

We thank the referee for a detailed and helpful report on the paper.

\label{lastpage} 
\end{document}